\documentclass[aip,reprint,superscriptaddress]{revtex4-1}

\usepackage{graphicx}
\usepackage{siunitx}
\usepackage{hyperref}
\usepackage{color}

\begin{document}

\newcommand*\mycommand[1]{\texttt{\emph{#1}}}

\author{Kathrin Ganzhorn}
\email{kathrin.ganzhorn@wmi.badw.de}
\affiliation{Walther-Mei\ss ner-Institut, Bayerische Akademie der Wissenschaften, 85748 Garching, Germany}
\affiliation{Physik-Department, Technische Universit{\"a}t M{\"u}nchen, 85748 Garching, Germany}

\author{Tobias Wimmer}
\affiliation{Walther-Mei\ss ner-Institut, Bayerische Akademie der Wissenschaften, 85748 Garching, Germany}
\affiliation{Physik-Department, Technische Universit{\"a}t M{\"u}nchen, 85748 Garching, Germany}

\author{Joel Cramer}
\affiliation{Institute of Physics, Johannes Gutenberg-University Mainz, 55099 Mainz, Germany}
\affiliation{Graduate School of Excellence Materials Science in Mainz, Staudinger Weg 9, 55128 Mainz, Germany}

\author{Richard Schlitz}
\affiliation{Walther-Mei\ss ner-Institut, Bayerische Akademie der Wissenschaften, 85748 Garching, Germany}
\affiliation{Physik-Department, Technische Universit{\"a}t M{\"u}nchen, 85748 Garching, Germany}
\affiliation{Institut f\"ur Festk\"orperphysik, Technische Universit\"at Dresden, 01062 Dresden, Germany}

\author{Stephan Gepr{\"a}gs}
\affiliation{Walther-Mei\ss ner-Institut, Bayerische Akademie der Wissenschaften, 85748 Garching, Germany}

\author{Gerhard Jakob}
\affiliation{Institute of Physics, Johannes Gutenberg-University Mainz, 55099 Mainz, Germany}

\author{Rudolf Gross}
\affiliation{Walther-Mei\ss ner-Institut, Bayerische Akademie der Wissenschaften, 85748 Garching, Germany}
\affiliation{Physik-Department, Technische Universit{\"a}t M{\"u}nchen, 85748 Garching, Germany}
\affiliation{Nanosystems Initiative Munich, 80799 Munich, Germany}

\author{Hans Huebl}
\affiliation{Walther-Mei\ss ner-Institut, Bayerische Akademie der Wissenschaften, 85748 Garching, Germany}
\affiliation{Physik-Department, Technische Universit{\"a}t M{\"u}nchen, 85748 Garching, Germany}
\affiliation{Nanosystems Initiative Munich, 80799 Munich, Germany}

\author{Mathias Kl{\"a}ui}
\affiliation{Institute of Physics, Johannes Gutenberg-University Mainz, 55099 Mainz, Germany}
\affiliation{Graduate School of Excellence Materials Science in Mainz, Staudinger Weg 9, 55128 Mainz, Germany}

\author{Sebastian T.B. Goennenwein}
\affiliation{Walther-Mei\ss ner-Institut, Bayerische Akademie der Wissenschaften, 85748 Garching, Germany}
\affiliation{Physik-Department, Technische Universit{\"a}t M{\"u}nchen, 85748 Garching, Germany}
\affiliation{Institut f\"ur Festk\"orperphysik, Technische Universit\"at Dresden, 01062 Dresden, Germany}
\affiliation{Nanosystems Initiative Munich, 80799 Munich, Germany}

\title{Temperature dependence of the non-local spin Seebeck effect in YIG/Pt nanostructures}

\begin{abstract}
We study the transport of thermally excited non-equilibrium magnons through the ferrimagnetic insulator YIG using two electrically isolated Pt strips as injector and detector. The diffusing magnons induce a non-local inverse spin Hall voltage in the detector corresponding to the so-called non-local spin Seebeck effect (SSE). We measure the non-local SSE as a function of temperature and strip separation. In experiments at room temperature we observe a sign change of the non-local SSE voltage at a characteristic strip separation $d_0$, in agreement with previous investigations. At lower temperatures however, we find a strong temperature dependence of $d_0$. This suggests that both the angular momentum transfer across the YIG/Pt interface as well as the transport mechanism of the magnons in YIG as a function of temperature must be taken into account to describe the non-local spin Seebeck effect.
\end{abstract}

\maketitle

\begin{figure}%
\includegraphics[width=\columnwidth]{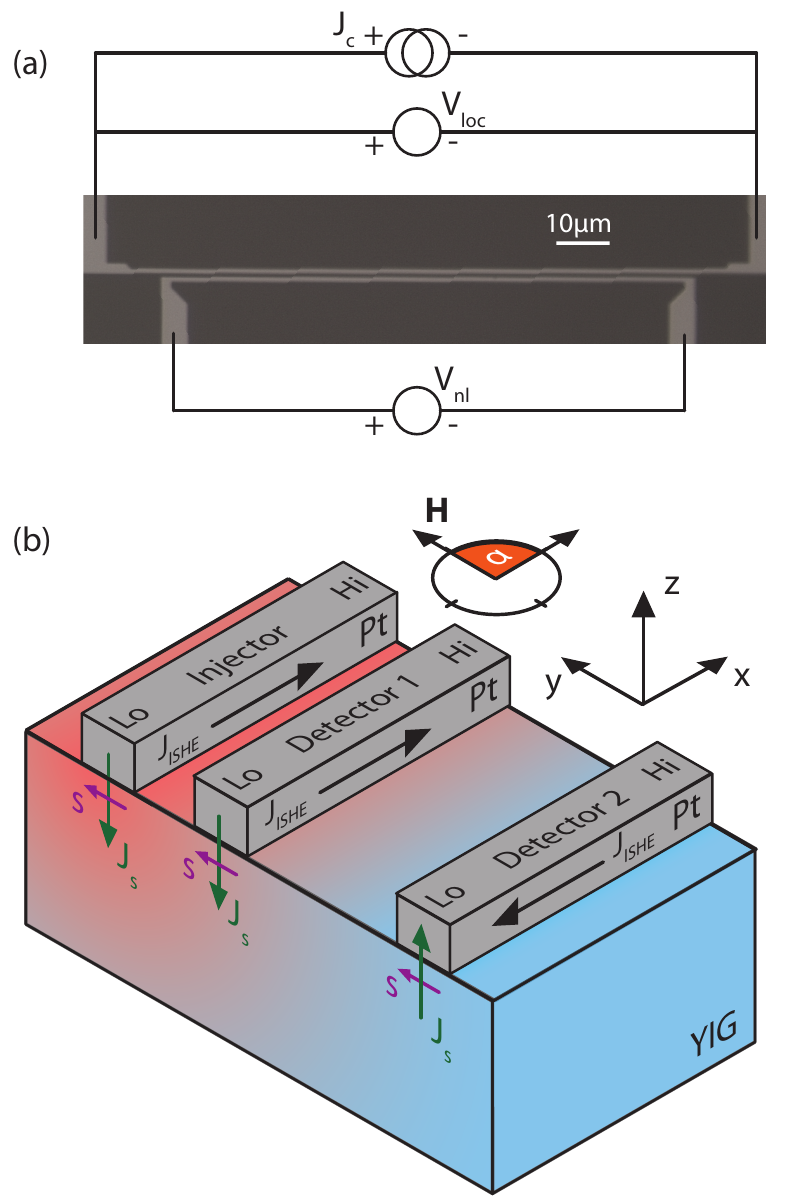}%
\caption{(a) Optical micrograph of a non-local nanostructure with two Pt strips (bright) on a YIG film (dark), including the electrical wiring. (b) Sketch of the YIG/Pt heterostructure: a dc charge current $\textbf{J}_c$ (not shown here) is applied to the injector strip (left) and the spin Seebeck signal is detected locally via the ISHE. The corresponding non-local thermal signal is measured along the detector strips for different strip separations. The color coding gives a qualitative profile of the magnon accumulation $\mu_m$ in the YIG film, where red corresponds to $\mu_m < 0$ (magnon depletion) and blue to a positive $\mu_m$ (magnon accumulation). In the short distance regime $\mu_m < 0$ at the injector and detector, such that the same sign is expected for local and non-local SSE. With increasing distance from the injector, $\mu_m$ and consequently the spin current across the interface as well as the detected ISHE voltage change sign.}%
\label{SSEmodel}%
\end{figure}

Magnons, the collective excitations in magnetically ordered systems, represent an attractive option for information transfer and processing. Using the ferrimagnetic insulator Yttrium Iron Garnet (YIG) as a model system, magnon-based information processing schemes have been put forward on the basis of coherently excited spin waves \cite{Chumak_review, Chumak_magnon_transistor, Khitun_logic}. Recent experiments in YIG/Pt heterostructures furthermore show that information can also be carried by incoherent non-equilibrium magnons \cite{CornelissenNat2015, Goennenwein2015} diffusing in YIG. This approach even allows for the implementation of logic operations within the magnetic system \cite{Ganzhorn_magnonlogic}. The non-equilibrium magnons can be excited and detected electrically via spin scattering mechanisms at the YIG/Pt interface. In a scheme referred to as magnon-mediated magnetoresistance (MMR), magnons are generated by driving a dc charge current through an injector Pt strip and detected as a non-local voltage in a second strip. The MMR effect has been studied as a function of the distance $d$ between injector and detector \cite{CornelissenNat2015}, temperature \cite{Goennenwein2015} and magnetic field magnitude and orientation \cite{CornelissenField}, allowing for the extraction of the length scales involved in the magnon diffusion process. In addition to the electrical injection, non-equilibrium magnons can also be generated thermally, via local Joule heating in the injector strip. The ensuing thermal non-local voltage is called non-local spin Seebeck effect in analogy to the well established local (longitudinal) spin Seebeck effect (SSE) \cite{SSEUchida}. While the microscopic mechanisms and in particular the relevant length scales for the SSE have been investigated in quite some detail \cite{Kehlberger_SSE, Guo_SSE_thickness}, the physics behind the non-local thermal signal is not well established. In this context, the non-local SSE has recently been studied at room temperature as a function of strip separation $d$ and YIG thickness \cite{Shan_thickness_dependence}. While at short distances $d$ the local and the non-local SSE signals have the same sign, for larger distances the non-local SSE amplitude is inverted, which was attributed to the profile of the non-equilibrium magnon distribution in the YIG film.

In this letter, we systematically study the non-local spin Seebeck effect in YIG/Pt nanostructures as a function of temperature and strip separation and find that the non-local SSE voltage changes sign at a characteristic strip separation $d_0$, which is strongly temperature dependent. We interpret our findings as evidence of a complex interplay between the temperature dependences of the interfacial transparency, i.e. angular momentum transfer across the YIG/Pt interface, and the diffusive properties of the thermally excited non-equilibrium magnons.

We investigate the non-local spin Seebeck effect in YIG/Pt bilayers fabricated and nano-patterned at the Walther-Mei\ss ner-Institut (sample series A) and at Johannes Gutenberg-University Mainz (sample series B). Series A was fabricated starting from a commercially available \SI{2}{\micro\meter} thick YIG film grown onto (111) oriented Gd$_3$Ga$_5$O$_{12}$ (GGG) via liquid phase epitaxy (LPE). After Piranha cleaning and annealing (see Ref.~\citenum{Puetter_APL} for details) to improve the interface quality, \SI{10}{\nano\meter} of Pt were deposited onto the YIG film using electron beam evaporation. A series of nanostructures consisting of 2 parallel Pt strips with length $l=\SI{100}{\micro\meter}$, width $w=\SI{500}{\nano\meter}$ and an edge-to-edge separation of $\SI{20}{\nano\meter}\leq d\leq \SI{10}{\micro\meter}$ were patterned using electron beam lithography followed by Ar ion etching. Series B was fabricated using a \SI{3.3}{\micro\meter} thick LPE-YIG film grown onto a GGG substrate as well. A series of strips with width $w=\SI{1}{\micro\meter}$ and edge-to-edge distance $\SI{100}{\nano\meter}\leq d\leq \SI{1.2}{\micro\meter}$ were patterned using electron beam lithography followed by a lift-off process with a Pt thickness of \SI{7.5}{\nano\meter}. An optical micrograph of one of these nanostructures is depicted in Fig.~\ref{SSEmodel} (a). 

In order to study the devices in series A and B, the samples were mounted in the variable temperature insert of a superconducting magnet cryostat ($\SI{10}{\kelvin} \leq T \leq\SI{300}{\kelvin}$). For series A an external magnetic field $\mu_0 H=\SI{1}{\tesla}$ was rotated in the thin film plane, while for series B the external magnetic field was applied along the $y$-direction and swept from $-\SI{250}{\milli\tesla}$ to $+\SI{250}{\milli\tesla}$ ($\alpha=\SI{90}{\degree}$, \SI{270}{\degree} in Fig.~\ref{SSEmodel} (b)). For local longitudinal spin Seebeck effect measurements in one single (injector) strip, we used the current heating method described in Ref.~\citenum{Schreier_iSSE}: a charge current $J_c=\SI{100}{\micro\ampere}$ is applied to the Pt strip along the $x$ direction using a Keithley 2400 source meter, inducing Joule heating in the normal metal. The ensuing temperature gradient across the Pt/YIG interface gives rise to the spin Seebeck effect and generates a spin current $\textbf{J}_s$ flowing across the interface, with the spin current spin polarization $\textbf{s}$ determined by the orientation of the magnetization $\textbf{M}$ in YIG. This spin current is accompanied by a charge current $J_\mathrm{ISHE}$ flowing along the $x$-direction in the Pt, as shown in Fig.~\ref{SSEmodel} (b). The voltage drop $V_\mathrm{loc}$, which includes the local SSE and the resistive response of the injector strip is recorded using a Keithley 2182 nanovoltmeter. Since the spin Seebeck effect is of thermal origin, the SSE voltage is proportional to the Joule heating power in the Pt and therefore independent of the heating current direction. Using the switching scheme of Ref.~\citenum{Schreier_iSSE}, we extract $V_\mathrm{therm,loc}= (V_\mathrm{loc}(+J_c)+V_\mathrm{loc}(-J_c))/2$, and thereby eliminate additional resistive effects such as the spin Hall magnetoresistance.

\begin{figure}
\includegraphics[width=\columnwidth]{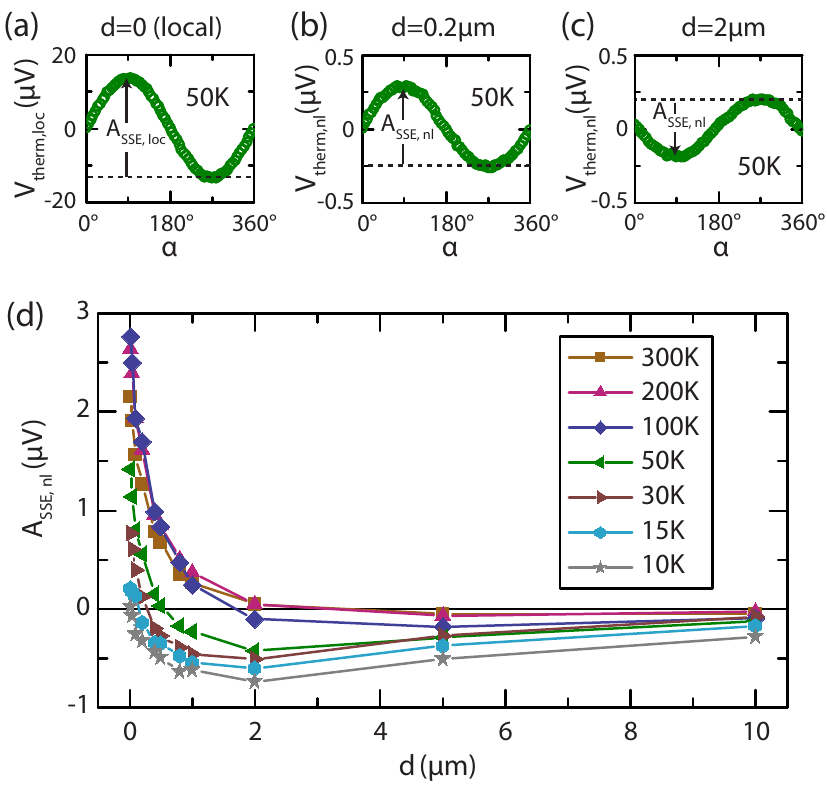}%
\caption{(a) Local spin Seebeck voltage detected at the injector strip at \SI{50}{\kelvin} as a function of the in-plane magnetic field orientation $\alpha$ with respect to the $x$ axis. (b), (c) Non-local thermal voltage detected at $T=\SI{50}{\kelvin}$ at the second strip for a strip separation of $d=\SI{200}{\nano\meter}$ and $\SI{2}{\micro\meter}$, respectively. (d) Non-local SSE amplitude $A_\mathrm{SSE,nl}$ extracted from in-plane field rotations at temperatures between $\SI{10}{\kelvin}$ and $\SI{300}{\kelvin}$ as a function of the injector-detector separation $d$.}%
\label{SSE_vs_d}%
\end{figure}

Figure \ref{SSE_vs_d} (a) shows $V_\mathrm{therm,loc}$ measured as a function of the magnetic field orientation $\alpha$ with respect to the $x$-axis at \SI{50}{\kelvin} for a device from series A. We observe the characteristic SSE dependence $V_\mathrm{therm,loc} \propto \mathrm{sin}(\alpha)$ yielding a positive amplitude $A_\mathrm{SSE,loc}=V_\mathrm{therm,loc}(\SI{90}{\degree})-V_\mathrm{therm,loc}(\SI{270}{\degree})$ of the local SSE, as expected in YIG/Pt heterostructures for this field configuration \cite{Schreier_SSE_sign}.

\begin{figure}%
\includegraphics[width=\columnwidth]{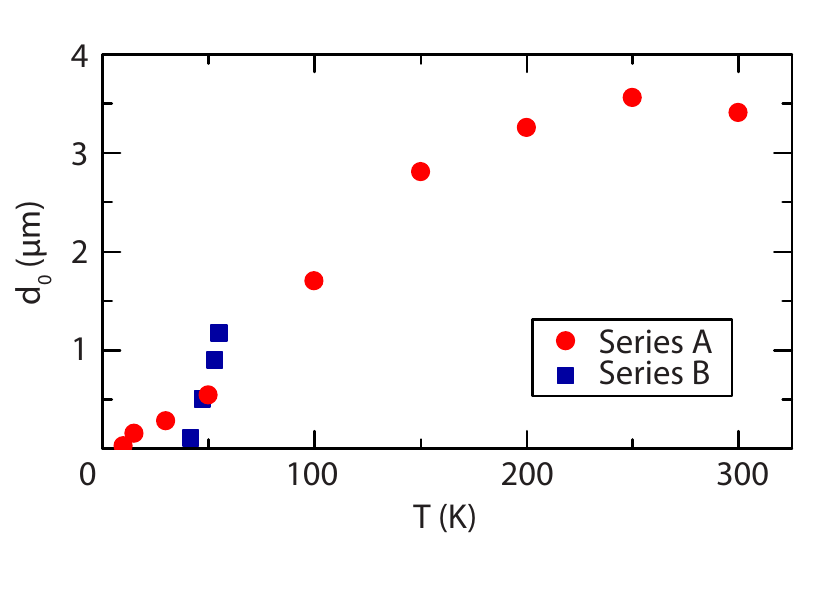}%
\caption{Temperature dependence of the critical strip separation $d_0$ at which the non-local SSE changes sign for sample series A (red) and B (blue).}%
\label{d0_vs_T}%
\end{figure}

Using an additional nanovoltmeter, we simultaneaously measure the voltage drop $V_\mathrm{nl}$ arising along the unbiased and electrically isolated second Pt strip. In analogy to the local thermal signal, the non-local thermal voltage is extracted as $V_\mathrm{therm,nl}= (V_\mathrm{nl}(+J_c)+V_\mathrm{nl}(-J_c))/2$ in order to distinguish it from resistive non-local effects such as the magnon mediated magnetoresistance \cite{Goennenwein2015}. $V_\mathrm{therm,nl}$ as a function of the external magnetic field orientation at \SI{50}{\kelvin} is depicted in Fig.~\ref{SSE_vs_d} (b) and (c) for strip separations of $d=\SI{200}{\nano\meter}$ and $\SI{2}{\micro\meter}$ (series A). In both devices, we observe a sin$(\alpha)$ dependence, with an amplitude $A_\mathrm{SSE,nl}$ about one order of magnitude smaller than for the local SSE. While the signal is positive for the $d=\SI{200}{\nano\meter}$ device, a negative $A_\mathrm{SSE,nl}$ is observed in the device with a larger injector-detector separation of $d=\SI{2}{\micro\meter}$. In order to confirm this sign change, we extract the amplitude of the non-local SSE measured at \SI{50}{\kelvin} in different devices with strip separations ranging from $\SI{20}{\nano\meter}$ to $\SI{10}{\micro\meter}$. The resulting data is shown in Fig.~\ref{SSE_vs_d} (d) as green symbols. Indeed, a sign change is observed at a strip separation $d_0 \approx \SI{560}{\nano\meter}$. Repeating these measurements as a function of temperature in the range between \SI{10}{\kelvin} and \SI{300}{\kelvin} yields the data compiled in Fig.~\ref{SSE_vs_d} (d). For all temperatures, a sign change in $A_\mathrm{SSE,nl}$ is observed as a function of the strip separation. Invariably, for small gaps the local and non-local SSE are both positive, but for large gaps the non-local SSE becomes negative. The experimental data in Fig.~\ref{SSE_vs_d} (d) show that the critical strip separation $d_0$, which is defined by $A_\mathrm{SSE}=0$, shifts to larger values as the temperature increases. The values $d_0$ extracted from Fig.~\ref{SSE_vs_d} for different temperatures are shown in Fig.~\ref{d0_vs_T} as red symbols for sample series A. With increasing temperature $d_0$ increases monotonically and seems to saturate around $T=\SI{200}{\kelvin}$. 

Similar experiments as a function of temperature and strip separation were performed on devices from series B and the critical strip separation extracted from these measurements is included in Fig.~\ref{d0_vs_T} as blue squares. While the temperature dependence is much steeper, a qualitatively similar increase of $d_0$ with temperature is observed in both series.

This characteristic sign change in the non-local SSE in YIG/Pt heterostructures above a particular separation $d_0$ has been previously observed by Shan et al. \cite{Shan_thickness_dependence} at room temperature and was attributed to the spatial profile of the non-equilibrium magnon accumulation $\mu_m$ in the YIG film, as shown schematically in Fig.~\ref{SSEmodel} (b). Magnons are thermally excited in the ferrimagnet due to Joule heating in the injector strip and diffuse vertically towards the GGG/YIG interface as well as laterally to the sample edges. According to the model proposed by Shan et al. \cite{Shan_thickness_dependence}, this leads to a depletion of magnons ($\mu_m < 0$, red in Fig.~\ref{SSEmodel} (b)) compared to the thermal equilibrium population beneath the injector. On the other hand, diffusing magnons accumulate further away from the injector, giving rise to $\mu_m > 0$ (blue in Fig.~\ref{SSEmodel} (b)). This model is applied to describe the increase of $d_0$ with increasing YIG thickness observed by Shan et al. \cite{Shan_thickness_dependence}: in contrast to phonons, the magnons cannot cross the YIG/GGG interface and accumulate there. As a consequence, $d_0$ (which marks the sign change of $\mu_m$) shifts to smaller values for thinner YIG films. Note that since the overall profile of $\mu_m$ is governed by diffusive magnon transport, the corresponding length scales can reach several \si{\micro\meter} \cite{Shan_thickness_dependence}. 
As shown in Fig.~\ref{SSEmodel} (b), the sign of $\mu_m$ determines the direction of the interfacial spin current $\textbf{J}_s$ at the detector, i.e. towards (away from) the YIG for negative (positive) $\mu_m$ at detector 1 (detector 2), and consequently governs the sign of the measured non-local ISHE voltage. Non-local SSE measurements as a function of the strip separation therefore allow us to map out the non-equilibrium magnon distribution in the YIG film. In particular the characteristic length $d_0$ for the sign change of $\mu_m$ can be determined.

In order to rationalize the measured temperature dependence of $d_0$, the parameters governing the angular momentum transfer across the YIG/Pt interface as well as the magnon diffusion process need to be analyzed. 
It has been shown that the transparency of the YIG/Pt interface, described by the effective spin-mixing conductance $g_s$, influences the magnon accumulation and hence the sign-reversal distance $d_0$ \cite{Shan_thickness_dependence}. For a fully opaque interface (obtained using an Al$_2$O$_3$ interlayer) which suppresses angular momentum backflow into the injector and therefore preserves a strong magnon depletion, an increase of the sign-reversal distance $d_0$ was observed \cite{Shan_thickness_dependence}. Previous measurements of the MMR effect in a YIG/Pt heterostructure  as a function of temperature have shown that the MMR signal decreases with decreasing temperature \cite{Goennenwein2015}, consistent with $g_s \propto T^{3/2}$ as predicted by theory \cite{ZhangPRB, CornelissenPRB94}. However, with this temperature dependence we expect a decreasing transparency of the YIG/Pt interface with decreasing temperature, leading to an increase of $d_0$ at low temperatures according to the model presented in Ref.~\citenum{Shan_thickness_dependence}. Since this is not consistent with our experimental observations depicted in Fig.~\ref{d0_vs_T}, the interface properties alone are not sufficient to describe the temperature dependence of the non-local SSE.

In addition to the interfacial transparency, the magnon diffusion length $\lambda_m$ and the magnon spin conductivity $\sigma_m$ determine the spatial distribution of the non-equilibrium magnons in YIG. We extracted the magnon diffusion length from temperature dependent MMR measurements conducted in the sample series A and found an increase of $\lambda_m$ with decreasing temperature by about a factor of 3 between \SI{300}{\kelvin} and \SI{50}{\kelvin}, following a 1/$T$ dependence. This is different from the temperature independent diffusion length reported by Cornelissen et al. \cite{CornelissenJUL2016}, who extracted $\lambda_m(T)=\mathrm{const.}$ together with a magnon spin conductivity $\sigma_m$ vanishing at low temperatures. While the detailed evolution of $\lambda_m$ and $\sigma_m$  with $T$ thus must be studied more systematically in future work, it is clear that these quantities depend on temperature. This implies that they can qualitatively impact the magnon diffusion process and consequently the non-local SSE. Indeed, the strong dependence of $d_0$ on the relative amplitudes of $g_s$, $\lambda_m$ and $\sigma_m$ at a fixed temperature has been demonstrated by Shan et al. using a one-dimensional analytical model for the spin Seebeck effect \cite{Shan_thickness_dependence}. 

Based on the available experimental data we can conclude that for a quantitative modeling of the non-local SSE the temperature dependence of both the angular momentum transfer across the YIG/Pt interface and the magnon diffusion in YIG must be taken into account. We furthermore note that additional effects due to phononic heat transport, known to be of importance for the local SSE, cannot fully be excluded at this point as a source of the non-local SSE-like signal. The local SSE originates from a finite difference of the effective temperature of the magnon and phonon subsystems in YIG close to the YIG/Pt interface, giving rise to a magnon spin current \cite{SSEXiao}. While this difference may also influence the thermal signal measured at the non-local detector strip, it was shown that the thermalization of the magnon and phonon subsystems takes place on a length scale $\lambda_\mathrm{mp}$ of the order of a few nm at room temperature \cite{Schreier_temp_profile, Boona_Heremans, Flipse_Peltier_PRL}, due to a very efficient (magnon conserving) magnon-phonon scattering. The effective temperature model described in Ref.~\citenum{SSEXiao} is therefore applicable mainly in the local limit, i.e.~close to the injector. The long-distance non-local limit however will be dominated by diffusing magnons and the magnon-phonon scattering which does not conserve the number of magnons and can reach a larger length scale $\lambda_\mathrm{m}$ of the order of several $\mu$m \cite{CornelissenPRB94}, as discussed above.

In summary, we have measured the non-local SSE in a YIG/Pt heterostructure as a function of injector-detector distance and temperature. The non-local SSE changes sign at a characteristic injector-detector separation $d_0$, confirming previous observations put forward by Shan et al. \cite{Shan_thickness_dependence}. We furthermore observe a decrease of the characteristic separation $d_0$ with decreasing temperature. Our results suggest a complex dependence of the non-local SSE on interfacial transparency, magnon diffusion properties as well as phonon heat transport.

KG acknowledges N. Vlietstra for discussions and JC thanks S. Kauschke for sample preparation. This work is financially supported by the Deutsche Forschungsgemeinschaft through the Priority Program Spin Caloric Transport (GO 944/4, GR 1132/18, KL 1811/7) and the SFB TRR 173 Spin+X, the Graduate School of Excellence Materials Science in Mainz (MAINZ) and EU projects (IFOX, NMP3-LA-2012 246102, INSPIN FP7-ICT-2013-X 612759).

%\bibliography{nonlocal_SSE_YIG}

\end{document}